\documentclass[prb,twocolumn,showpacs]{revtex4}

\usepackage{graphicx}
\usepackage{amsmath}
\usepackage{amssymb}

\def\g{\gamma}
\def\e{\varepsilon}
\def\d{\delta}

\def\t{\tau}
\def\n{\nu}
\def\o{\omega}
\def\p{\varphi}
\def\r{\rho}
\def\s{\sigma}

\def\S{\Sigma}
\def\G{\Gamma}

\def\O{\Omega}

\def\x{\xi}

\def\bk{{\bf k}}
\def\bK{{\bf K}}

\def\bn{{\bf n}}

\def\bp{{\bf p}}
\def\bn{{\bf n}}

\def\be{\begin{equation}}
\def\ee{\end{equation}}
\def\bea{\begin{eqnarray}}
\def\eea{\end{eqnarray}}
\def\nn{\nonumber}
\def\lb{\label}

\begin{document}
\title{Electric bias control on impurity effects in bigraphene}
\author{Y.G. Pogorelov,$^1$ M.C. Santos,$^2$ V.M. Loktev$^{3,4}$}
\affiliation{$^1$IFIMUP-IN, Departamento de F\'{\i}sica, Universidade do Porto, Rua do Campo
Alegre 687, Porto 4169-007, Portugal,\\
$^2$Departamento de F\'{\i}sica, Universidade de Coimbra, R. Larga, Coimbra 3004-535, Portugal,\\
$^3$ Bogolyubov Institute for Theoretical Physics, NAN of Ukraine, 14b Metrologichna str.,
Kiev 03680, Ukraine\\
$^4$ National Technical University of Ukraine “KPI”, Pr. Peremogy 37, Kiev 03056, Ukraine}

\begin{abstract}
Formation of localized impurity levels within the band gap in bigraphene under applied electric
field is considered and the conditions for their collectivization at finite impurity concentration
are established. It is shown that a qualitative restructuring of quasiparticle spectrum within the
initial band gap and then specific metal-insulator phase transitions are possible for such disordered
system at given impurity perturbation potential and concentration, such processes being effectively
controlled by variation of the electric field bias. Since these effects can be expected at low enough
impurity concentrations and accessible applied voltages, being stable enough thermically, they can be
promising for practical applications in nanoelectronics devices.
\end{abstract}

\maketitle\

\section{Introduction}
\lb{int}
Between various derivatives from the basic graphene system \cite{geim,wall,sem}, a special interest
is attributed to its bilayer combination, the bigraphene \cite{novo}. This interest is mainly due to
the important possibility to realize here a case of semiconductor with controllable band gap through
the application of an electric field \cite{zhang, castro, ando}. It should be noted that similar
crystalline structure of two planes with hexagonal lattices is now recognized for a whole family of
materials, either really fabricated or theoretically predicted. Besides the two known modifications of
bigraphene itself, the Bernal (or A-B) structure \cite{bernal} and its alternative, A-A structure
\cite{tao}, there exist also the bilayers of silicene, the silicon analog to graphene \cite{liu}, the
bilayers of boron nitride \cite{rib} or its bilayered combination with graphene \cite{slaw}, the bilayered
chalcogenides of transition metals (pure or alloyed) \cite{ramas}, etc. However, the most reliable
structure for external tuning and the simplest for theoretical study is seen in the Bernal-stacked
bigraphene, hence chosen here as the basic host system for studying impurity effects.  By an analogy
with the known effects of dopants in common semiconductor systems, this opens a possibility for specific
localized impurity levels to exist within the host spectrum gap \cite{nils,feher}, like those known for
common donor and acceptor levels by impurities in conventional semiconductors with fixed band structure
\cite{sze,shklov}. Next, it is known that, at high enough impurity concentration, an intensive interaction
between the localized impurity states related to these levels can take place, and this can essentially
modify the band spectrum near the gap edge \cite{iva,zhang1}, giving rise to specific narrow energy ranges
of band-like states near impurity levels (called impurity bands) and even producing a phase transition from
insulating to metallic state \cite{mott} with important practical applications. Then the case of in-gap
impurity states in bigraphene could provide an even more flexible field of electronic properties due to the
possibility of continuous control on band gap, thus permitting controllable phase transitions. Such situation
was recognized long ago in some magnetic crystals with impurities where the magnetic excitation spectra and
so the observable properties are controllable by applied magnetic field \cite{ilp}. To the authors' knowledge,
such a possibility has not been known before for fermionic systems, and it could open interesting perspectives
for future nanoelectronics.

The paper is organized as follows. In Sec. \ref{ham}, the second quantization Hamiltonian is defined for a
Bernal-stacked biased bigraphene (free from impurities) and the related matrix representation for Green
functions (GF) is built, giving rise to its 4-subband electronic spectrum. Sec. \ref{imp} introduces
the model impurity perturbation and studies formation of impurity levels and the conditions for their possible
development into impurity bands, based on specific forms of self-energy matrices present in the GF matrices.
A more detailed study on such impurity bands, including the estimates for characteristic mobility edges between
their ranges of band-like and localized states, is done in Sec. \ref{band}. Then the possibility for specific
metal-insulator phase transitions in doped bigraphene under variation of external electric field bias (at fixed
impurity concentration) and the resulting transport effects are considered in Sec. \ref{mit}. The final Sec.
\ref{conc} discusses the main conclusions and suggestions for practical applications of the described impurity
effects.

\section{Bigraphene under applied field}
\lb{ham}
As is well known, the relevant electronic dynamics of a graphene sheet are generated by the carbon $sp^3$
orbitals (whose energy level can be chosen as the energy reference) in the simplest approximation of single
hopping parameter $t$ between nearest neighbor carbons from different sublattices at distance $a$ in the
honeycomb lattice \cite{wall}. The bigraphene structure, furthermore,  involves the interlayer hopping $t_z$
by vertical links between nearest neighbors from different sublattices (for Bernal stacking) shown in Fig.
\ref{fig1}. With an account taken of an electric bias $V = eEd$ between the layers (with the electron charge
$e$, the applied electric field $E$, and the interlayer spacing $d$), this defines the tight-binding (Fourier
transformed) Hamiltonian $4\times 4$ matrix:
\begin{figure}
 \includegraphics[width=8cm]{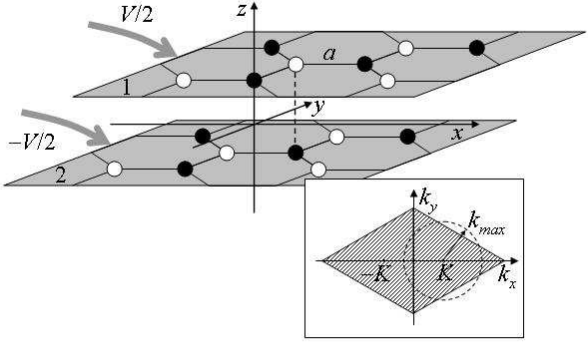}\\
  \caption{Schematic of Bernal-stacked bigraphene under applied electric bias $V$. The A- and B-type sites in
  each plane are indicated by black and white circles respectively, the solid and dashed lines indicate the
  in-plane $t$ and interplane $t_z$ links. Inset: the Brillouin zone in $\bk$-plane with two Dirac points
  $\pm\bK$ and an equivalent circle of radius $k_{max} = \sqrt{K/a}$.}
\lb{fig1}
\end{figure}
\be \hat H_\bk =  \left(\begin{array}{cccc}
                                            V/2 & \g_\bk & 0 & t_z \\
                                            \g_\bk^\ast & V/2 & 0 & 0 \\
                                            0 & 0 & -V/2 & \g_\bk \\
                                            t_z & 0 & \g_\bk^\ast & -V/2 \\
                                          \end{array}
                                        \right).
   \lb{eq1}
     \ee
Here the wave vector $\bk$ lies in the first Brillouin zone (see inset in Fig. \ref{fig1}) and the in-plane
dispersion follows from the sums $\g_\bk = t\sum_{\boldsymbol \d}{\rm e}^{i\bk\cdot{\boldsymbol \d}}$ over
nearest neighbor vectors $\boldsymbol \d$ of the honeycomb lattice, suitably approximated as $\g_\bk \approx
\x_\bk {\rm e}^{i\p_\bk}$ with $\x_\bk = \hbar v_{\rm F}|\bk - \bK|$, the Fermi velocity $v_{\rm F} =
3ta/2\hbar$, and $\p_\bk = \arctan k_y/(k_x - K_x)$ near the Dirac points $\bK =\pm (4\pi/3\sqrt 3 a,0)$ (the
relevant range of $|\bk - \bK| \sim K t_z/W$ is really narrow, since $t_z$ is weak besides the total bandwidth
$W$, see below). This matrix enters the second-quantization Hamiltonian (in absence of impurity perturbation):
\be
 H_0 = \sum_\bk \psi_\bk^\dagger \hat H_\bk \psi_\bk,
\lb{eq2}
\ee
where the spinors $\psi_\bk^\dagger = \left(a_{1\bk}^\dagger,b_{1\bk}^\dagger,a_{2\bk}^\dagger,b_{2\bk}^\dagger\right)$
are made of Fourier transformed second quantization operators $a_{j\bk} = N^{-1}\sum_\bn a_{j\bn} {\rm e}^{i\bk\cdot\bn}$
and $b_{j\bk} = N^{-1}\sum_\bn b_{j\bn}{\rm e}^{i\bk\cdot\bn}$ with the on-site operators $a_{j\bn}$ and $b_{j\bn}$
for A- and B-type sites respectively from $n$th unit cell in $j$(= 1,2)-th layer and $N$ is the number of cells in
a layer. Generally, the energy spectrum is defined through the matrix of Fourier transformed two-time GFs \cite{bb,eco}
$\hat G_\bk =  \langle\langle \psi_{\bk}|\psi_{\bk}^\dagger\rangle\rangle$, as solutions of the dispersion equation:
\be
 {\rm Re\, det \,}\hat G_\bk^{-1} = 0.
 \lb{eq3}
  \ee
Thus, for the non-perturbed system by Eq. \ref{eq2} the GF matrix reads $\hat G_\bk^{(0)} = (\e - \hat H_\bk)^{-1}$
and, after diagonalization of $\hat H_\bk$ in spinor indices, its dispersion near the Dirac points is suitably
expressed through the radial variable $\x_\bk \equiv \x$. It includes two positive energy subbands:
\be
 \e_{\n}(\x) = \sqrt{\frac{t_z^2}2 + \frac{V^2}4 + \x^2 - (-1)^\n  \sqrt{\frac{t_z^4}4 + \x^2 \left(t_z^2 +
  V^2\right)}},
  \lb{eq4}
   \ee
the "external" ($\n = 1$) and "internal" ($\n = 2$) ones, and their negative energy counterparts, as shown in Fig.
\ref{fig2}a. The most relevant feature of this spectrum is the bias-controlled energy gap between the extrema
$\pm\e_g = \pm V/(2\sqrt{1 + (V/t_z)^2})$ of two internal subbands, attained along a circle around each Dirac point
(the so-called "mexican hat") whose radius in the $\x$-variable is $\x_0 = \sqrt{\e_g^2 + V^2/4}$ .

The physical characteristics of this system follow from the GF matrix as, for instance, the density of states (DOS)
by electronic quasiparticles:
\be
 \r(\e) = \frac 1 {\pi} {\rm Im\,Tr}\, \hat G(\e),
  \lb{eq5}
   \ee
where $\hat G(\e) = (2N)^{-1} \sum_{\bk} \hat G_\bk(\e)$ is  the local GF matrix, and its imaginary part for exact
band spectrum results as usually from infinitesimal imaginary shift of energy argument, $\e - i 0$ \cite{eco}. In
what follows, the sum in $\bk$ over triangular halves of the Brillouin zone is approximated by the $\x$- integration:
\[\frac 1 {2 N} \sum_{\bk}f_\bk(\e) \approx \frac 2{W^2} \int_0^W f(\x,\e)\x d\x,\]
over two equivalent circles around Dirac points (inset in Fig. \ref{fig1}) of the $\x$-radius $W = \hbar v_{\rm F}
k_{max}$ (where $k_{max} = \sqrt{K/a}$, see inset in Fig. \ref{fig1}), well justified at low energies, $|\e| \ll W$.
For the pure bigraphene system by Eq. \ref{eq1}, the result for Eq. \ref{eq5} is generated by the explicit diagonal
elements of non-perturbed local GF matrix \cite{nils}:
\bea G_{11}^{(0)} & \approx & 2\frac{\e - \e_2}{W^2}\left[\frac{\e\e_2}{\d^2 }\left(\pi - \arctan\frac{\d^2}
{\e^2 + \e_2^2}\right) + \ln\frac\g W\right],\nn\\
G_{22}^{(0)} & = & G_{11}^{(0)}(\e) - t_z^2\frac{\e + \e_2}{W^2\d^2}\left(\pi - \arctan\frac{\d^2}{\e^2 + 
\e_2^2}\right),
\lb{eq6}
\eea
where
\bea
\d^2(\e) & = & \sqrt{\left(t_z^2 + V^2\right)\left(\e_g^2 - \e^2\right)},\nn\\
 \g^2(\e) & = & \sqrt{\left(\e^2 - \e_1^2\right)\left(\e^2 - \e_2^2\right)}.\nn
\eea
These elements reveal the inverse square root divergences at the gap edges $\pm \e_g$ (of Im $G$ beyond the gap
and of Re $G$ within the gap), also note the finite steps of Im $G$ at the limiting energies $\e_{1,2} \equiv
\e_{1,2}(0)$ of the two subbands. The resting diagonal elements are simply $G_{33}^{(0)}(\e) = -G_{22}^{(0)}(-\e)$
and $G_{44}^{(0)}(\e) = -G_{11}^{(0)}(-\e)$, so that finally DOS is a function of $\e^2$, as shown in Fig
\ref{fig2}b in agreement with the known previous calculations \cite{ando}. It presents BCS-like divergences near
$\e_g^2$, finite steps at $\e_{1,2}^2$, and gets to coincidence with the linear DOS for monolayer graphene
\cite{wall,sem} beyond $\e_1^2$, due to joint (non-linear) contributions from both subbands.
\begin{figure}
 \includegraphics[width=8cm]{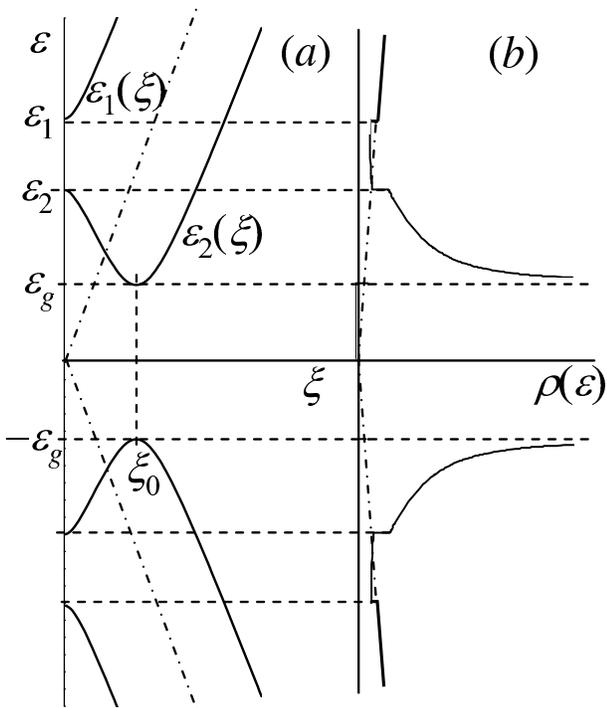}\\
  \caption{a) Dispersion laws for the bilayer in Fig. \ref{fig1} {\it vs} the radial variable $\x$ near a Dirac
  point, given by Eq. \ref{eq4} at the choice of $V = 2t_z$, the dash-dotted line indicate the Dirac dispersion
  for monolayered   graphene. b) DOS for this choice, the dash-dotted line marks the linear DOS for monolayered
  graphene.}
\label{fig2}
 \end{figure}

Within the gap, only real parts of $G_{jj}(\e)$ are non-zero, and their divergences near the gap edges are crucial
for appearance, under the effect of localized impurity perturbations, of in-gap localized levels and related
collective states which is the main focus for the analysis below.

\section{Impurity levels and impurity subbands}
\lb{imp}
As was recognized from experimental studies on graphene systems \cite{ara}, they can contain a variety of defects,
ranging from topological ones (vacancies, dislocations, edges, boundaries, etc.) to impurity adatoms (or some
functional groups) near one of planes and in-plane substitutes or interstitials. This provides a doping of charge
carriers (of both signs) into these systems as well as scattering of carriers on impurity potentials and possibly
formation of localized (or resonance) impurity states on such potentials. The latter must be characterized by some
model parameters within the common tight-binding approximation and the simplest case is the Lifshitz model only
involving the on-site perturbation potential $U$, identical for all impurity sites randomly distributed among the
lattice sites \cite{lif}. This model with moderate $U$ value (comparable with the graphene bandwith $W$) looks more
adequate to the case of substitutional impurities in graphene, than its unitary limit, $U \gg W$ in Ref. \cite{dahal}
or the alternative choice of Anderson model \cite{and} with random perturbations at each lattice site in Ref.
\cite{sheng}. Another alternative is the Anderson hybride (or {\it s-d}) model \cite{and1} with two parameters, the
impurity binding energy and its coupling to the host excitations, and its use for the so called deep impurity levels
in semiconductors is known to result in formation of the above mentioned impurity bands and related phase transitions.
However, such perturbation model when introduced into the framework of 4-component host spectrum of Sec. \ref{ham}
could make the most important treatment of interactions between impurities and of impurity band coherence technically
unfeasible. This determines our choice for the Lifshitz model (though known to provide less freedom for impurity
bands formation than the {\it s-d} model). By similar reasons, we do not consider the long-range impurity potentials
as for Coulomb \cite{biswas,pereira} or screened-Coulomb \cite{bena} centers.

Let us build the perturbation Hamiltonian by Lifshitz impurities on certain impurity sites. In accordance with
the composition of $\psi$-spinors, the A- and B-sites from first plane can be referred to the types $j = 1,2$
respectively and those from second plane to $j = 3,4$, then $\bp_j$ denote the defect sites of $j$th type with
relative concentrations $c_j = \sum_{\bp_j}N^{-1}$ such that the total impurity concentration $\sum_j c_j = c \ll 1$.
Then the sought Hamiltonian in terms of local Fermi operators reads
\bea
 H_1 & = & U \left(\sum_{\bp_1} a_{1\bp_1}^{\dagger} a_{1\bp_1} + \sum_{\bp_2} b_{1\bp_2}^{\dagger}
  b_{1\bp_2}\right.\nn\\
& + & \left.\sum_{\bp_3} a_{2\bp_3}^{\dagger} a_{2\bp_3} + \sum_{\bp_4} b_{2\bp_4}^{\dagger}
  b_{2\bp_4}\right),
   \lb{eq7}
    \eea
or, in terms of $\psi$-spinors by Eq. \ref{eq2}, it takes the form of scattering operator:
\be
 H_1 = \frac 1{N} \sum_{j,\bp_j}\sum_{\bk,\bk'}{\rm e}^{i\left(\bk' - \bk\right)\cdot\bp_j} \psi_\bk^\dagger
  \hat U_j \psi_{\bk'}.
    \lb{eq8}
      \ee
where the diagonal matrix $ \hat U_j$ has a single non-zero element $U$ at the $jj$ site. Considering now the
Hamiltonian in presence of impurities $H_0 + H_1$ and following a similar routine to Refs. \cite{pls,pls1}, we
arrive at solutions for the GF matrix in two specific forms adequate for two alternative types of excitation
states in a disordered system \cite{mott,lif}, the band-like (extended) states and localized states. Thus, the
first of these types is better described by the so-called fully renormalized representation (FRR) of GF \cite{ilp}:
\be
 \hat G_\bk = \left[\left((\hat G_\bk^{(0)}\right)^{-1} - \hat\S_\bk\right]^{-1}
  \lb{eq9}
   \ee
where the self-energy matrix is additive in different types of impurity centers:  $ \hat\S_\bk = \sum_j
\hat\S_{j,\bk}$, with the partial matrices given by the related FRR group expansions (GE's) in complexes of
impurity centers (of the same $j$-type, involved in multiple scattering processes):
\bea
 \hat\S_{j,\bk} & = & c_j\hat T_j \left[1 + c_j {\sum_{\bn \neq 0}}\left({\rm e}^{-i\bk\cdot\bn} \hat A_{j,\bn}
  + \hat A_{j,\bn} \hat A_{j,-\bn}\right)\right.\nn\\
& \times & \left. \left(1 - \hat A_{j,\bn} \hat A_{j,-\bn}\right)^{-1} + \dots\right].
  \lb{eq10}
   \eea
Each T-matrix $\hat T_j = \hat U_j\left(1 - \hat G \hat U_j\right)^{-1}$ describes all the scatterings on a
single impurity center of $j$th type, and the next to unity term in r.h.s. of Eq. \ref{eq10} accounts for
scatterings on pairs of $j$-impurities through the matrices $\hat A_{j,\bn} = \hat T_j(2N)^{-1}\sum_{\bk'
\neq \bk}\hat G_{\bk'} {\rm e}^{i\bk'\cdot\bn}$ of indirect interaction (via band-like excitations) in such
pairs at separation $\bn$. Notice the excluded quasimomentum $\bk$ (for given $\hat \S_\bk$) in this sum, also
the FRR GE excludes coinciding quasimomenta in all the multiple sums for products of interaction matrices
\cite{ilp}. The omitted terms in Eq. \ref{eq10} relate to all scattering processes in groups of three and more
impurities, and their general structure can be found in similarity with the known group integrals from the
Ursell-Mayer classical theory of non-ideal gases.

Otherwise, for the range of localized states, the non-renormalized representation (NRR):
\be
 \hat G_\bk = \hat G_\bk^{(0)} - \hat G_\bk^{(0)}\hat\S\hat G_\bk^{(0)},
  \lb{eq11}
   \ee
is more adequate. Here the respective NRR self-energy matrix $\hat\S = \sum_j \hat\S_j$ has a similar structure to
the FRR one by Eq. \ref{eq10} but with the NRR matrices $\hat T_j^{(0)} = \hat U_j(1 - \hat G^{(0)} \hat U_j)^{-1}$,
$\hat G^{(0)} =(2N)^{-1}\sum_{\bk}\hat G_\bk^{(0)}$, and with no restrictions in all the quasimomentum sums for the
products of NRR interaction matrices $\hat A_{j,\bn}^{(0)} = \hat T_j^{(0)}(2N)^{-1} \sum_{\bk}\hat G_{\bk}^{(0)}
{\rm e}^{i\bk\cdot\bn}$ (that are only present in their even combinations $\hat A_{j,\bn}^{(0)} \hat A_{j,-\bn}^{(0)}$).
\begin{figure}
 \includegraphics[width=9.5cm]{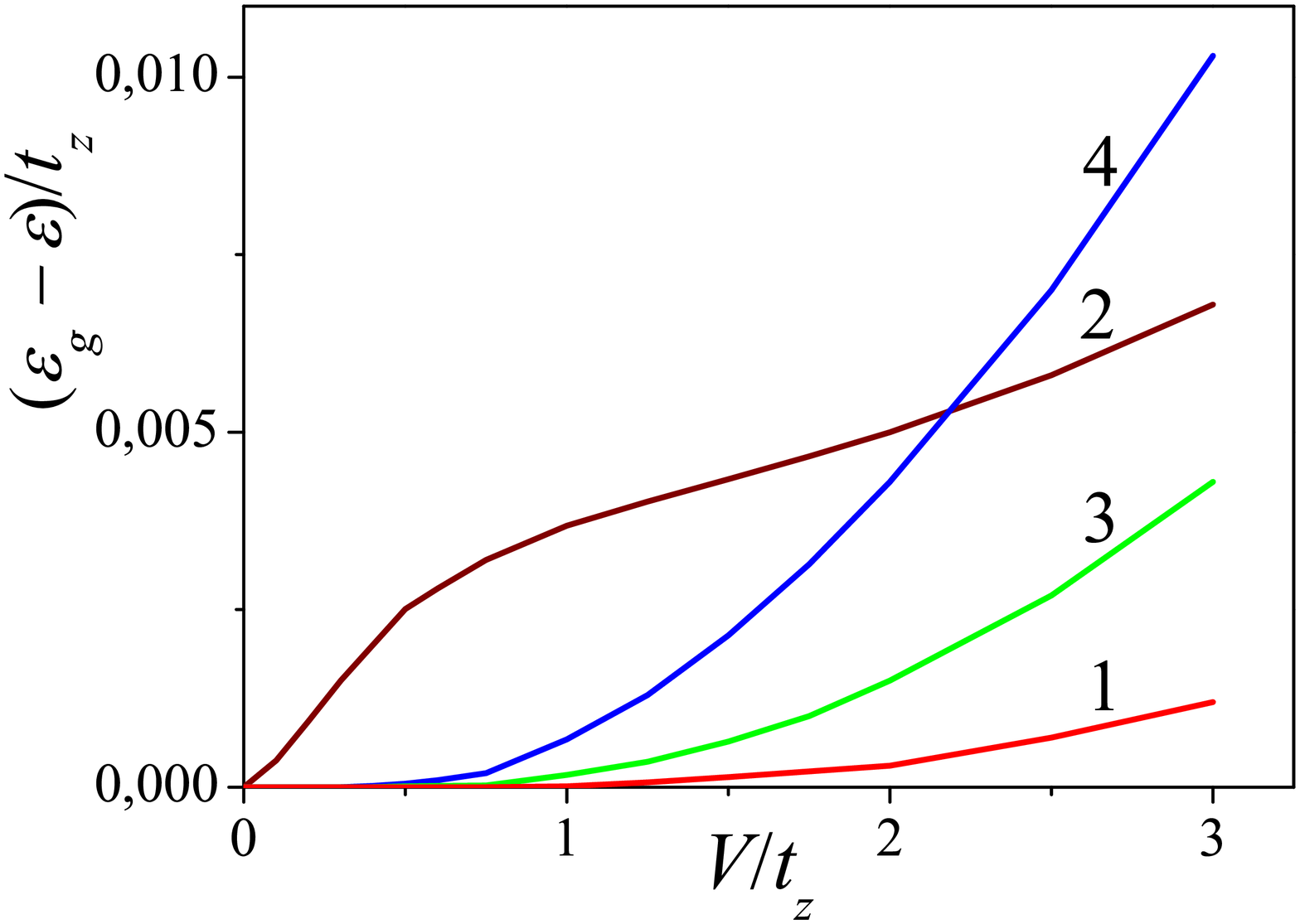}\\
  \caption{Separations of the in-gap impurity levels  $\e^{(j)}$ from the gap edge as functions of the applied bias
  $V$ (all   in   units of $t_z$, the curves being labeled by the $j$ numbers). Note the different behavior of
  $\e^{(1,2)}$ levels from   that of $\e^{(3,4)}$ ones and the resulting interchange of the deepest levels at the
  bias value $V_{cr} \approx 2.2t_z$ (see also the text).}
 \lb{fig3}
  \end{figure}
The best known effect of local perturbations consists in emergence of localized energy levels within the band gap and
those were already indicated for impurities in bigraphene \cite{nils,feher}. As known from general theory \cite{lif,
ilp}, such levels are most pronounced at sufficiently low concentration of impurities (so that their indirect
interactions can be neglected) and given by the poles of T-matrices. In the present case, the matrices $\hat T_j^{(0)}$
give rise to four different local levels $\e^{(j)}$ within the band gap, and their locations depend on the magnitude and
sign of perturbation parameter $U$ (like the known situations in common doped semiconductors \cite{sze,shklov}) but yet
on the applied field $V$ (as a specifics of doped bigraphene). The positions of four impurity levels $\e^{(j)}$ by each
type of impurity center are the roots of related Lifshitz equations:
\be
 U G_{jj}^{(0)}(\e^{(j)}) = 1,
  \lb{eq12}
   \ee
so that choosing for definiteness $U = -W/2$ and using Eq. \ref{eq6} provides their dependence on the applied
bias $V$ as shown in Fig. \ref{fig3} (for their relative separations from the gap edge). It is seen that generally
they stay rather shallow at growing $V$, but with a notable difference between the levels $\e^{(1,2)}$ (by
impurities in the positive biased layer) and $\e^{(3,4)}$ (by those in the negative biased one). In particular,
a specific interchange of the deepest levels occurs in this course, from $\e^{(2)}$ to $\e^{(4)}$, at $V_{cr}
\approx 2.2t_z$ for given $U $. This feature was not indicated in the former analysis of the same model in Ref.
\cite{nils}, where  $\e^{(4)}$ was considered to remain the deepest level at all $V$ values. Also, it can be
noted that for the commonly used value of $t_z \approx 0.35$ eV this interchange bias would amount to $V_{cr}
\approx 0.77$ eV, well above the experimentally realized (to the moment) $V$ values of up to $\approx 0.36$ eV
\cite{zhang}. Thus, the much stronger separation of the $\e^{(2)}$ level at lower bias voltages could be of
significant practical importance.

The well known property of localized states by shallow energy levels is their long effective radius \cite{ilp},
also indicated for impurities in biased bigraphene \cite{nils}, defining intensive interactions between them
already at their very low concentrations. Such interactions were shown to allow, at certain conditions,
collectivization of impurity states to form specific band-like states within narrow energy bands (called
impurity bands) around the initial localized levels \cite{iva}. As will be seen below, this effect is possible
as well in the present case of multiple localized levels, where the most essential specifics is their joint
participation in forming the lowest impurity subband of much stronger dispersion than in higher lying subbands
(if those are permitted).

Formally, in similarity to the non-perturbed case, the band spectrum for the disordered system can be evaluated
from the dispersion equation, Eq. \ref{eq3}, with the GF matrix by FRR Eqs. \ref{eq9},\ref{eq10}. Of course, if
treated rigorously, it presents a tremendous problem of developing infinite sequence of renormalization procedures
in all possible terms of the corresponding GE, and there is no reasonable hope for its exact solution. The popular
way to avoid this problem is restriction of the full self-energy to its self-consistent T-matrix form (that is,
neglecting all the interactions between impurities) known as the coherent potential approximation (CPA)
\cite{soven} and it was suggested for studies of disorder effects both in monolayer graphene \cite{carva} and
bigraphene \cite{nils}. Although this approach describes certain impurity bands, it is known to fail just in
reproducing the observed (when available) narrow dispersion of these bands. On the other hand, validity of the
simplest NRF, Eq. \ref{eq11}, is only limited to the energy ranges of fully localized states.

A more consistent approach can be suggested using partial renormalizations of the full self-energy in Eq.
\ref{eq10}, first substituting there the NRR T-matrix and interaction matrices and then subsequently introducing
such approximate self-energies into the next generations of GF and interaction matrices, checking convergence of
the obtained GE's in order not to extend the renormalizations to irrelevant GE terms. Namely, it is reasonable to
define the $l$th generation GF matrix $\hat G_\bk^{(l)}$ by an analog to Eq. \ref{eq9} with the respective
self-energy $\hat \S_\bk^{(l)}$ by an analog to Eq. \ref{eq10} but containing T-matrices $\hat T^{(l-1)}$ and
interaction matrices $\hat A_\bn^{(l-1)}$ built with use of the preceding generation $\hat G_\bk^{(l-1)}$ matrices.
This algorithm leads to the true FRR at $l \to \infty$. However, even its first non-trivial $l = 1$ approximation
can be reasonable for the band-like energy ranges where the true FRR GE converges.
\begin{figure}
   \includegraphics[width=8cm]{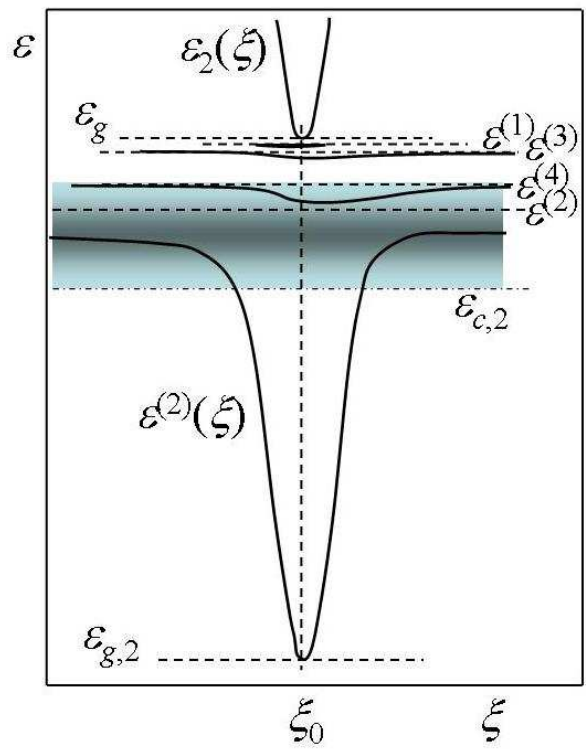}\\
  \caption{Formation of impurity subbands near the impurity levels by the solutions of Eq. \ref{eq3} in the 1st
  step   of   renormalization (see text) for the case of Fig. \ref{fig3} at $V = 2t_z$. The most dispersive
  $\e^{(2)} (\x)$   subband   extends well beyond the shadowed vicinity of the $\e^{(2)}$ level, that delimits
  the range of localized   states down to the respective mobility edge $\e_{c,2}$. Similar localized areas around
  $\e^{(4,3,1)}$ (not shown here) would continuously extend this range up to above the gap edge $\e_g$.}
 \lb{fig4}
  \end{figure}

Then, in the first step of this routine, the formal solutions of Eq. \ref{eq10} with the self-energies in the NRR
T-matrix approximation, $\hat \S_{j,\bk} \approx c_j\hat T_j^{(0)}$, display four narrow subbands near four impurity
levels $\e^{(j)}$, besides the four broad principal bands $\pm\e_\n(\x)$ (here only slightly modified compared to
Eq. \ref{eq4}). An example of such modified spectrum (at a natural choice of equal partial concentrations $c_j = c$
and taking the total impurity concentration $4c = 0.01$) for the cases of Fig. \ref{fig3} is shown in Fig. \ref{fig4}.
The lowest impurity subband, conventionally denoted here as $\e^{(2)}(\x)$ by its proximity to the lowest $\e^{(2)}$
level, is seen to strongly dominate in its dispersion over all the resting ones, and the direct analysis of Eq.
\ref{eq10} shows that this domination is due to the above mentioned constructive interplay between all $\e^{(j)}$.

Note that all the impurity subbands in this approximation produce BCS-like divergences in DOS, as well near the
levels $\e^{(j)}$ as near subbands terminations. However, since quasimomentum is not true quantum number in a
disordered system \cite{lif}, the analysis of its real energy spectrum, especially for the in-gap states, should
also take account of the damping $\G_j(\x)$ of each $\e^{(j)}(\x)$ state, resulting from ${\rm Im\,}\S_j$. Hence
one can consider these states Bloch-like (or conducting) only if the Ioffe-Regel-Mott (IRM) criterion \cite{ir,mott}
is fulfilled or the GE, Eq. \ref{eq10}, is convergent at related energies. Otherwise they should pertain to the
localized type. As will be seen, all the formal DOS singularities fall within the localized energy ranges and so
are effectively broadened.

The mentioned criteria permit to estimate also the division points between band-like and localized ranges known as
Mott's mobility edges \cite{mott}. Of course, such mobility edges can be found near the limits of both principal
and impurity bands, but our main focus here will be on the most dispersive impurity band, like $\e^{(2)}(\x)$ in
the above example, as the most interesting object for practical purposes. Finally, a certain special value
$V_{\rm A}$ of bias control (at given impurity concentrations $c_j$ and perturbation parameter $U$) can be indicated,
such that mobility edges from both sides of a conducting impurity band will merge. This  collapse of impurity band
will manifest a kind of Anderson transition \cite{and} in a disordered system, realized in a controllable way at
$V \to V_{\rm A}$.

It should be underlined that all these fundamental features of the energy spectrum in a disordered system are fully
ignored when the  narrow impurity bands are treated using the CPA approximation (as, e.g., in Refs. \cite{nils,carva})
beyond its known validity checks \cite{el}.

\section{Conditions for existence of impurity subbands}
\lb{band}
As known from studies on many disordered systems where a localized impurity level $\e_{imp}$ near an edge $\e_g$
of pure crystal energy band  can give rise, at high enough impurity concentration, to a specific impurity band
$\e_{imp}(\bk)$ \cite{ilp}, the latter is restricted by the general IRM criterion:
\be
\bk \cdot {\bf \nabla}_\bk \e_{imp}(\bk) \gg \G_{imp}(\e_{imp}(\bk)),
\lb{eq13}
\ee
where the linewidth $\G_{imp}(\e)$ of a Bloch-like state with quasi-momentum $\bk$ and energy $\e$ is defined as
the imaginary part of the corresponding self-energy. For the present multiband system, this criterion should be
formulated for each of $\e^{(j)}(\x)$ subbands by expanding the general determinant from Eq. \ref{eq3} near a given
energy $\e$ in a complex form: det $\hat G_\bk^{-1} \approx \left[\e - \e^{(j)}(\x) + i\G_j(\e)\right]\O_j (\e)$, to
obtain the corresponding linewidth $\G_j(\e)$ (aside a certain factor $\O_j(\e)$ of energy to cube dimension).

In the adopted Lifshitz model, each partial T-matrix $\hat T_j$ (regardless of its renormalization) has a single
non-zero element at the $jj$ site (alike $\hat U_j$ itself): $T_j = U/(1 - UG_{jj})$. In the above suggested first
step renormalization, we have Im$T_j^{(0)} = 0$ for $\e$ within the bandgap. Then the imaginary part of related
self-energy function $\S_j^{(1)}$ is here only due to the GE terms next to unity in Eq. \ref{eq10}, dominated by
the pair term once GE is convergent. It can be also shown that the most relevant contribution to Im$\S_j(\e)$ comes
from the $jj$th matrix element of the GE pair term while those from its other elements (though generally non-zero)
are strongly reduced by the quantum interference effects. This contribution:
\be
 B_j(\e) = {\rm Im}\sum_{n > a}\frac{A_{j,\bn}^{(0)}A_{j,-\bn}^{(0)}}{1 - A_{j,\bn}^{(0)}A_{j,-\bn}^{(0)}},
  \lb{eq14}
   \ee
can be obtained from the residues at zeroes of the denominator, using the explicit spatial behavior of scalar
interaction functions (see Appendix \ref{app} for details):
\bea
A_{j,\bn}^{(0)}(\e) & = & \frac{T_j^{(0)}(\e)}{2N}\sum_\bk {\rm e}^{i\bk\cdot\bn}\left(G^{(0)}_\bk\right)_{jj}\nn\\
& \approx & \sqrt{\frac{r_{j,\e}}n}{\rm e}^{-n/r_\e}\sin\frac{n}{r_0}\cos\bK\cdot\bn,
 \lb{eq15}
  \eea
where the characteristic scales are:
\[ r_{j,\e} = r_0\left(\pi\frac{\e_g - \e^{(j)}}{\e - \e^{(j)}}\right)^2, \quad r_\e = r_0\frac{\x_0^2}{\d^2},
\quad r_0 = \frac{\hbar v_{\rm F}}{\x_0}.\]
A similar behavior with two oscillating factors in effective inter-impurity interactions was previously indicated
for the impurity states within superconducting gap in ferropnictides \cite{pls} where a faster cosine factor had
Fermi wavelength. But, the present case is simplified by the specific symmetry of $\bK$ in the Brillouin zone, so
that $\cos^2\bK\cdot\bn$ for all separations $\bn$ between lattice sites of the same $j$th type only takes the
values $\s$ = 1 and 1/4 (with respective weights $p_\s$ = 1/3 and 2/3) whose contributions can be then simply
added up in Eq. \ref{eq14}. These partial contributions are obtained by subsequent integrations \cite{pls}, first
over the poles of fast oscillating sine and then over its residues with the slow envelope function $F_{j,n,\s}^2 =
\s r_{j,\e}{\rm e}^{-2n/r_\e}/n$:
\bea
 B_j & = & \sum_{\s}p_\s{\rm Im}\sum_{n > a} \frac{F_{j,n,\s}^2\sin^2(n/r_0)}{1 - F_{j,n,\s}^2\sin^2(n/r_0)}\nn\\
 &\approx &\sum_{\s}\frac{4\pi p_\s}{\sqrt 3 a^2}\int_a^{r_{max}}\frac{r dr}{\sqrt{F_{j,r,\s}^2 -1}},
  \lb{eq16}
   \eea
where $r_{max}$ corresponds to $F_{j,r_{max}} = 1$. The latter integration is simplified within the energy range of:
\be
 \e^{(j)} - \e \gg (\e_g - \e^{(j)})^{5/4}/\e_g^{1/4},
  \lb{eq17}
   \ee
where $r_{j,\e} \ll r_\e$ so that the exponential factor in Eq. \ref{eq15} remains approximately unity for all
distances $r < r_{max} \approx r_{j,\e}$. In this approximation, the explicit result for the most dispersive
subband reads:
\be B_2(\e) = \frac{7\pi}{64} \left(\frac{r_{2,\e}}a\right)^2,
 \lb{eq18}
  \ee
with the prefactor resulted precisely from weighting of $\s$ values. Then the above suggested expansion of
det $\hat G_\bk^{-1}$ for $\e$ closer to $\e^{(2)}$ than to other $\e^{(j)}$ (so that all $\S_j$ except $\S_2$
can be neglected) provides the linewidth:
\be
 \G_2(\e) \approx c^2(\e^{(2)} - \e)B_2(\e),
  \lb{eq19}
   \ee
valid until $\e^{(2)} - \e \lesssim  \e_g - \e^{(2)}$. For $\e$ yet farther from $\e^{(2)}$, we have $r_{j,\e}
< r_0$ so that $B_2(\e)$ vanishes and finite $\G_2$ values can only result from the higher order GE terms (if we
exclude, of course, all other possible relaxation processes, such as thermal phonons, electron-electron collisions,
etc.).  From Eq. \ref{eq19}, the IRM criterion is reduced to the inequality:
\[c B_2(\e) \ll 1\]
(agreeing with the GE convergence) and, supposing Eq. \ref{eq17} valid, this criterion permits to estimate the
mobility edge separation from the $\e^{(2)}$ level:
\be
 \e^{(2)} - \e_{c,2} \sim c^{1/4}\sqrt{\frac W {2\x_0}}\left(\e_g - \e^{(2)}\right).
  \lb{eq20}
   \ee
All the states with energies closer to $\e^{(2)}$ than $\e_{c,2}$ are localized on various clusters of 2nd type
impurity centers. The first conclusion from the estimate, Eq. \ref{eq20}, is that existence of the impurity
subband itself is only assured if its bandwidth $\approx \e^{(2)} - \e_{g,2}$ surpasses the width of localized
range around $\e^{(2)}$. This is fulfilled when the total impurity concentration exceeds the critical value:
\bea
c_{cr} & \sim & \left(\frac{t_z}{W}\right)^{8/3}\left(\frac {|U|} W\right)^{4/3}\left(\frac V W\right)^{2/3}\nn\\
& \times & \frac{(t_z + \sqrt{t_z^2 + V^2})(2t_z^2 + V^2)}{(t_z^2 + V^2)^{2/3}t_z^{5/3}},
 \lb{eq21}
  \eea
(it is obtained approximating Eq. \ref{eq6} only to its diverging terms). Smallness of this expression is mainly
due to its first three essential factors of interlayer coupling, impurity perturbation, and applied bias, while
the last factor stays almost constant for all realistic (not too high) $V$. Thus, for the sample choice of
parameters, $W = 20t_z$, $|U| = 10t_z$, and $V = 2t_z$, we obtain $c_{cr} \sim 1.8\cdot 10^{-5}$. Then for the
example of $c = 0.01$ chosen in Fig. \ref{fig3}, the mobility edge $\e_{c,2}$ extends from $\e^{(2)}$ to about
0.38 of the distance $\e_g - \e^{(2)}$ while the dispersion of $\e^{(2)}(\x)$ subband is about four times bigger
(see Fig. \ref{fig4}). Finally, from comparison of Eqs. \ref{eq20} and \ref{eq17} it follows that for $c > c_{cr}$
the latter vicinity always occurs within the localized range and so the exponential factor in Eq. \ref{eq15} cannot
influence the above obtained estimates.
\begin{figure}
\includegraphics[width=10cm]{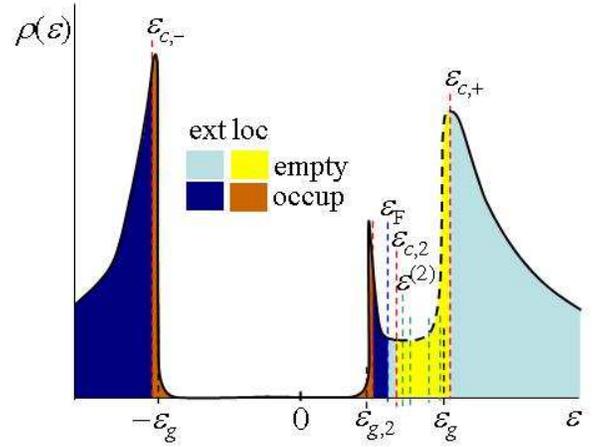}\\
  \caption{Schematics of extended (ext) and localized (loc) ranges in the energy spectrum of bigraphene with impurities
  for the situation alike that in   Fig. \ref{fig4}. Note the position of the Fermi level (separating occupied and empty
  states) with respect to the mobility edges (separating ext and loc states); the narrow impurity band only emerges below
  the lowest impurity level $\e^{(2)}$ (the labels for higher $\e^{(j)}$ are omitted) while the localized states
  fill the whole range from $\e_{c,2}$ up to $\e_{c,+}$ (see in text).}
    \lb{fig5}
     \end{figure}
In summary, only the most dispersive impurity subband by the lowest impurity level can be considered to really emerge
beyond its mobility edge, with its main specifics in anomalously strong variation of the lifetimes $\t(\e)$ along very
narrow energy intervals. As to other formal solutions of Eq. \ref{eq3} (analyzed with inclusion of the resting $B_j$),
they are mostly invalidated within the common overlapped range of higher laying localized levels, that extends up to
$\e_{c,+}$, the mobility edge of the upper main band. The states in this area can be only characterized by their DOS.
Though the latter function can not be directly found here from the above defined GE's, Eqs. \ref{eq9}, \ref{eq11}, it
can be plausibly expected to vary slowly until matching to the peak near $\e_g$ (Fig. \ref{fig5}), so that the total
number of states $\int_{-W}^W\rho(\e)d\e = 4$ is kept.

Similarly, some finer details of the energy spectrum can be also determined, such as, for instance, the rest of mobility
edges: $\e_{c,\pm}$, that define the broadened edges of main subbands, and that near the extremum $\e_{g,2} \approx
\e^{(2)}(\x_0)$ of the impurity subband (see Fig. \ref{fig5}). At last, the case of low impurity concentration, $c < c_{cr}$,
can be also included when there is no impurity band within the gap but the localized levels $\e^{(j)}$ turn to be separately
resolved. However all these data are not so relevant for our main practical purpose below and so left beyond the present
scope. Nevertheless, the presented results give an important extension and diversification of the general scenario of
collective restructuring of spectra of elementary excitations in crystals with impurities under external fields \cite{ilp}.

\section{Biased metal-insulator transitions and their observable effects}
\lb{mit}
Now we can pass to the important processes of electric transport in the system with the above described band spectrum.
For simplicity, this consideration is restricted to the case of zero temperature and the main attention is paid to the
position of Fermi level $\e_{\rm F}$ and to the lifetime $\t_{\rm F}$ of Fermi states under the applied bias control
$V$ at given parameters of impurity perturbations $c$ and $U$. The basic condition for the Fermi level:
\be
 2\int_{-\infty}^{\e_{\rm F}}\r(\e)d\e = 1 + c',
  \lb{eq22}
   \ee
defines its shift from the zero energy position in the unperturbed system, in order to accomodate the total of $c'$
extra carriers per unit cell (brought by impurities themselves and possibly by some external sources). This generally
requires knowledge of DOS functions for all the impurity subbands (besides almost non-perturbed ones for main subbands).
But our main interest here is in finding a possibility for $\e_{\rm F}$ to be located within the most dispersive impurity
band $\e^{(2)}(\x)$, so we focus on the related DOS, especially in proximity to this band termination $\e_{g,2}$ (Fig.
\ref{fig5}). An important simplification of this task is obtained by noting that for this energy range all the
self-energies $\S_j$ in Eq. \ref{eq10} can be taken as constants, small enough compared to the gap width, thus the
solutions of the dispersion equation, Eq. \ref{eq3}, almost reproduce here the non-perturbed  $\e_2(\x)$ band within
accuracy to a constant shift of its edge from $\e_g$ to $\e_{g,2}$ (see Fig. \ref{fig4}), just due to the common effect
of all $\S_j$. The resulting DOS function:
\be
 \r_2(\e) \approx \frac{2\e}{W^2}\frac{t_z^2 + V^2}{\d^2},
  \lb{eq23}
   \ee
at $0 < \e - \e_{g,2} \lesssim \e^{(2)} - \e_{g,2}$ defines from Eq. \ref{eq21} the Fermi level $\e_{\rm F}$ position
by the equation
\be
  c' \approx \left(\frac 2W\right)^2\sqrt{(t_z^2 + V^2)(\e_g^2 - \e_{\rm F}^2)}.
   \lb{eq24}
    \ee
Let $ c'_{max}$ be the maximum permitted amount of carriers such that $\e_{\rm F}$ stays within the conducting range.
Then, for the case of Fig. \ref{fig4}, this value results $ c'_{max} \approx 4\cdot 10^{-3}$, that is, somewhat lower
than the proper impurity concentration $c = 10^{-2}$ in this case. Nevertheless, conduction through the impurity band
can be realized if $c'$ is brought below the indicated limit of $ c'_{max}$, e.g., by external compensation of a part
of charge carriers \cite{shklov}. Once this is assured, one can then strongly change the conductivity by raising the
applied $V$, since the localized range width by Eq. \ref{eq19} grows with bias faster than $\propto V^{2/3}$ against
the almost bias insensitive (at $V \lesssim V_{cr}$, see Fig. \ref{fig3}) width of the impurity band, while the Fermi
level $\e_{\rm F}$ goes to the band edge $\e_{g,2}$ slower than $\propto V^{-2}$.  Then the faster advancing mobility
edge $\e_{c,2}$ will finally cross $\e_{\rm F}$ at some bias $V_{M-I}$, giving rise to a Mott metal-insulator transition
and vanishing conductivity. Thus, for the proposed choice of $U = -W/2 = -10t_z$ and $c' = 3\cdot 10^{-3}$, we obtain
$V_{M-I} \approx 0.87$ eV. In this course, at $V \to V_{M-I}$, conductivity can vary by orders of magnitude, when we
drive the Fermi inverse lifetime $\t_{\rm F}^{-1} \sim \G_2(\e_{\rm F})/ \hbar$ close to divergence, under very tiny
variations (say, some meV) of bias. This indicates a tremendous potentiality of such type of doped semiconducting
systems in comparison with traditional materials.

Besides their evident field transistor applications, critical effects by the biased Mott transition can be also
expected in other observable properties of this doped system, for instance, in its optical response at the frequency
$\o_{i,b} \approx \left(\e^{(2)} + \e_g\right)/\hbar$ of transition from the top of occupied $-\e_2(\x)$ band and the
Fermi states of impurity $\e^{(2)}(\x)$ band (like the case formerly considered by the authors for doped superconducting
iron pnictides \cite{pls1}) that can be switched on and off by tiny variation of the bias.

At last, with further growing bias, the collapse of upper and lower mobility edges within the impurity band and the
aforementioned Anderson transition to fully localized in-gap spectrum can be realized. From Eq. \ref{eq20} at $V \lesssim
t_z$, this bias value estimates as $V_{\rm A} \sim c^{3/2}W^7|U|^{-2}t_z^{-4}$, though this analytic expression only applies
(at moderate $|U|$) for as low impurity concentrations as $c \lesssim 10^{-5}$ . However a numerical analysis with use of
full Eq. \ref{eq6} shows that $V_{\rm A}$ remains attainable up to $c \sim 10^{-2}$ as well. This transition can also
produce observable effects; in this case the collapse of narrow impurity band would lead to a dramatic drop of the
plasmonic resonance frequency \cite{zay}.

\section{Discussion and conclusions}
\lb{conc}
The above main conclusion about the possibility to attain extensive control on electrical conduction through very slight
variations of applied potential implies, of course, many additional factors to be taken into account. They can be indicated
both from the theoretical and practical sides. Thus, the used theoretical approach is restricted to a simple model of
impurity perturbation by a single on-site parameter, and some elaboration of it could be done involving, for instance,
perturbations of hopping parameters. These kinds of analyses are known for traditional semiconductors with impurities and
also have demonstrated possibility for similar bands of collective states to be formed near localized impurity levels at
high enough impurity concentrations. Notably, for those more common materials, it was just the Lifshitz perturbation model
that presented the biggest theoretical problems for such effects, for instance, by leading to unrealistically high values
of critical concentration, of order of unity or even more (unlike that in Eq. \ref{eq19}). This permits to expect that
modifications of the present Lifshitz model, as in Ref. \cite{feher} for single impurity at gapless spectrum, or using the
Anderson hybrid model \cite{and1} as in Ref. \cite{mkhit} (provided all the technical aspects be assured) will not change
essentially the physical behavior of the system. On the other hand, there are yet many properties of this simple model that
can be further studied, for instance, the possibilities to realize multiple conducting impurity subbands and subsequent
processes of multiple switching between them, including, e.g., optical transitions under electrical biasing. Of course, a
more realistic approach should also take account of topological defects (see beginning of Sec. \ref{imp}) as well as the
above mentioned Coulomb interactions, thermal effects, etc. Generally, this would require the impurity band structure to
exceed a certain "background" relaxation level, that could be achieved by varying either the impurity sort (that is, $U$
parameter) and concentration or/and the applied bias $V$. Finally, similar impurity multiband effects can be also sought
in other atomically multilayered systems, such as those mentioned in the Introduction, where a special focus might be put
on tuned bandgap in silicene bilayers (yet wider than in bigraphene \cite{liu}) or even on single layers of buckled silicene
or germanene \cite{ni}.

As to the practical issues, first of all, rather strict conditions on fabrication of the basic doped bilayered system are
in order, perhaps mainly aimed to minimize all the "foreign" defects {\it vs} the chosen dopants, but the next requirement
to keep the levels of dopants (and possibly their compensating species) within to fractions of percent should not be a real
problem for modern nanoelectronics. A special attention is also required for precise control and manipulation of the applied
bias $V$, particularly, in exploring possibilities to realize its near-critical and super-critical regimes, like those
indicated in the above analysis. Finally, the practical arrangement of an experimental transistor-type setup based on the
suggested conductivity control by tiny impurity subbands would perhaps require some specific technical solutions. However,
they do not look too difficult to be found in the available engineering depository. Thus, a fair hope exists for this
theoretical proposal to be realized in a practical device.

In conclusion, the effects of localized on-site perturbations by rather disperse impurities on bilayered graphene system
under the applied electrical bias between the layers are theoretically considered using the Green function techniques
adapted for a multiband electronic system, demonstrating the conditions for different types of localized impurity energy
levels to appear within the bias-induced bandgap in the electronic spectrum of this system and then extension of these
levels into specific narrow energy bands at impurity concentration surpassing certain characteristic values. The analysis
on these processes demonstrated their similarities to those known from literature on various crystalline materials with
impurities. Also, some specifics of the present system were shown in considerable bias dependences of impurity bands and
of critical concentrations for their formation. These dependences can be further treated to provide some specific phase
diagrams in variables "bias-concentration" as it took place in antiferromagnetic insulators where such diagrams in variables
"magnetic field-concentration" were quite informative \cite{ilp}. A practical application of the described electronic band
structure is suggested in a form of highly sensitive bias control on the system's conductivity through the impurity subband
when brought close to a regime of bias-controlled Mott's metal-insulator transition.

\section*{Acknowledgements}

Y.G.P. is grateful for the support of this work from Portuguese FCT project PTDC/FIS/120055/2010 and M.C.S. is indebted
to the support from FCT project PTDC/QUI-QUI/117439/2010. V.M.L ackhowledges partial supports from the European FP7 program
by SIMTECH 246937 Grant, by STCU Grant No 5716-2 "Development of Graphene Technologies" and by the Special Program of
Fundamental Researches of NAS of Ukraine.

\appendix \section{}\lb{app}

In calculation of the interaction function, Eq. \ref{eq15}, the essential task consists in the integration:
\bea \frac{1}{2N}\sum_\bk {\rm e}^{i\bk\cdot\bn}\left(G^{(0)}_\bk\right)_{jj} & & =  \frac{2\cos\bK\cdot\bn}
 {W^2}\nn\\
\times  \int_0^W J_0\left(\frac{\x n}{\hbar v_{\rm F}}\right) & & \frac{(N_j(\e) - \x^2)\x d\x}{\left(\x^2 -
 \x_1^2\right)\left(\x^2 - \x_2^2\right)},
\lb{eq24}
 \eea
where $J_0$ is the zeroth order Bessel function, $\x_{1,2}^2 = \e^2 +  \e_2^2 \pm \d^2(\e)$ are the complex poles
of det$\hat G_\bk^{(0)}$ in $\x$-variable and all $|N_j(\e)| \sim \e_g^2$ (as seen from Eq. \ref{eq6}). Since this
integral is fast converging after $\x \gtrsim \e_g$, its upper limit can be safely extended to infinity.  Then, after
expanding the factor besides the Bessel function in simple fractions:
\be
\frac{N_j(\e,\x)}{\left(\x^2 - \x_1^2\right)\left(\x^2 - \x_2^2\right)} = \frac{N_j(\e) - \x_1^2}{\x^2 - \x_1^2} -
 \frac{N_j(\e) - \x_2^2}{\x^2 - \x_2^2},
  \lb{eq25}
   \ee
the Hankel-Nicholson integration formula can be applied to each of them:
\be
 \int_0^\infty \frac{J_0(x)x dx}{x^2 + z^2} = K_0(z),
  \lb{eq26}
   \ee
with the zeroth order MacDonald function $K_0$, valid for complex $z$ if Re$z > 0$ (Ref. \cite{abst}). The $z$-arguments
related to the terms in Eq. \ref{eq25}, can be defined as $z_{1,2}^2 = -\x_{1,2}^2(n/\hbar v_{\rm F})^2$ and the above
requirement will read Re$\sqrt{-\x_{1,2}^2} > 0$. For the relevant energy range $0 < \e_g - \e \ll \e_g$, we can use the
choices $\sqrt{-\x_{1,2}^2} = \sqrt{\sqrt{\g^2(\e) - \e^2 - \e_2^2}} \mp i\sqrt{\sqrt{\g^2(\e) + \e^2 + \e_2^2}}$. At
last, for relevant distances $n \gtrsim r_0$, the resulting $K_0(z_{1,2})$ have big enough arguments, $|z_{1,2}| =
|n\x_{1,2}/ \hbar v_{\rm F}| \gtrsim 1$, to use their asymptotics: $K_0(z_1) \approx -\sqrt{2/\pi z_1}{\rm e}^{-z_1}$
and $K_0(z_2) \approx \sqrt{2/\pi z_2}{\rm e}^{-z_2}$. Then, taking account of all prefactors besides these expressions,
present in Eqs. \ref{eq15} and \ref{eq25}, we arrive at the final result of Eq. \ref{eq15}.

\end{document}